\documentclass[aps,                
               prc,                
               showpacs,           
               notitlepage,
               onecolumn,          
               10pt,               
               twoside,            
               floatfix,           
               superscriptaddress] 
               {revtex4-1}

\usepackage{mathptmx}
\usepackage[T1]{fontenc}
\usepackage{geometry}         
\usepackage{fancyhdr}         
\usepackage{graphicx}         
\usepackage[svgnames]{xcolor} 
\usepackage{hyperref}         
\usepackage{amsmath}          
\usepackage{amssymb}          
\usepackage{amsfonts}         
\usepackage{amstext}          
\usepackage{textcomp}         
\usepackage{slashed}

\hypersetup{colorlinks=true,linkcolor=Blue,citecolor=Blue,urlcolor=Blue}

\DeclareMathAlphabet{\mathcal}{OMS}{cmsy}{m}{n}

\newcommand{\bs}[1]{p\!\!\!\!\! p \!\!\!\!\! p\!\!\!\!\! p}
\newcommand{\bsp}[1]{p\!\!\!\!\! p \!\!\!\!\! p\!\!\!\!\! p}
\newcommand{\bsq}[1]{q\!\!\!\!\! q}
\newcommand{\bsA}[1]{A\!\!\!\!\! A}
\newcommand{\bsP}[1]{P\!\!\!\!\! P}
\newcommand{\bsAA}[1]{A\!\!\!\!\!\!\! A}
\newcommand{\bsa}[1]{a\!\!\!\!\! a}
\newcommand{\bsu}[1]{u\!\!\!\!\! u}
\newcommand{\bsbt}[1]{\beta\!\!\!\!\!\! \beta \!\!\!\!\!\! \beta \!\!\!\!\!\! \beta}

\numberwithin{equation}{section}                                

\fancypagestyle{firststyle}
{%
  \fancyhf{}
  \fancyhead[C]{PHYSICAL REVIEW C {\bf 87}, 034912 (2013)} 
  \fancyfoot[L]{\small 0556-2813/2013/87(1)/034912(25)} 
  \fancyfoot[C]{\small 034912-\thepage} 
  \fancyfoot[R]{\small \textcopyright 2013 American Physical Society} 
}

\begin{document}

\title{\texorpdfstring{\begin{minipage}[c]{\textwidth}\centering The quark susceptibility in a
generalized dynamical quasiparticle model \end{minipage}}{Quark Susceptibilities: DQPM meets lQCD}}

\author{H.~Berrehrah}
\email{berrehrah@fias.uni-frankfurt.de}
\affiliation{\begin{minipage}[c]{\textwidth}Frankfurt Institute for Advanced Studies, Johann Wolfgang Goethe Universit\"at, Ruth-Moufang-Strasse 1, 60438 Frankfurt am Main, Germany \end{minipage}\\ \vspace{0.5mm}}
\affiliation{\begin{minipage}[c]{\textwidth}Institut for Theoretical Physics, Johann Wolfgang Goethe Universität, Max-von-Laue-Str. 1, 60438 Frankfurt am Main, Germany \end{minipage} \\ \vspace{0.5mm}}

\author{W.~Cassing}
\email{wolfgang.cassing@theo.physik.uni-giessen.de}
\affiliation{\begin{minipage}[c]{\textwidth}Institut für Theoretische Physik, Universit\"at Giessen,
35392 Giessen, Germany\end{minipage} \\\vspace{0.5mm}
\vspace{2.5mm}}

\author{E.~Bratkovskaya}
\email{brat@th.physik.uni-frankfurt.de}
\affiliation{\begin{minipage}[c]{\textwidth}Frankfurt Institute for Advanced Studies, Johann Wolfgang Goethe Universit\"at, Ruth-Moufang-Strasse 1, 60438 Frankfurt am Main, Germany \end{minipage}\\ \vspace{0.5mm}}
\affiliation{\begin{minipage}[c]{\textwidth}Institut for Theoretical Physics, Johann Wolfgang Goethe Universität, Max-von-Laue-Str. 1, 60438 Frankfurt am Main, Germany \end{minipage} \\ \vspace{0.5mm}}

\author{Th.~Steinert}
\email{thorsten.steinert@theo.physik.uni-giessen.de}
\affiliation{\begin{minipage}[c]{\textwidth}Institut für Theoretische Physik, Universit\"at Giessen,
35392 Giessen, Germany\end{minipage} \\\vspace{0.5mm} 
\vspace{2.5mm}}


\begin{abstract}

The quark susceptibility $\chi_q$ at zero and finite quark chemical
potential provides a critical benchmark to determine the
quark-gluon-plasma (QGP) degrees of freedom in relation to the
results from lattice QCD (lQCD) in addition to the equation of state
and transport coefficients. Here we extend the familiar
dynamical-quasiparticle model (DQPM) to partonic propagators that
explicitly depend on the three-momentum with respect to the partonic
medium at rest in order to match perturbative QCD (pQCD) at high
momenta. Within the extended dynamical-quasi-particle model
(DQPM$^*$) we reproduce simultaneously the lQCD results for the
quark number density and susceptibility and the QGP pressure  at
zero and finite (but small) chemical potential $\mu_q$. The shear
viscosity $\eta$ and the electric conductivity $\sigma_e$ from the
extended quasiparticle model (DQPM$^*$) also turn out in close
agreement with lattice results for $\mu_q$ =0. The DQPM$^*$,
furthermore, allows to evaluate the momentum $p$, temperature $T$
and chemical potential $\mu_q$ dependencies of the partonic degrees
of freedom also for larger $\mu_q$ which are mandatory for transport
studies of heavy-ion collisions in the regime 5 GeV $< \sqrt{s_{NN}}
<$ 10 GeV.
\end{abstract}


\maketitle

\section{Introduction}
\label{Introduction}

The thermodynamic properties of the quark-gluon plasma (QGP)--as
produced in relativistic heavy-ion collisions--is well determined
within lattice QCD (lQCD) calculations at vanishing quark chemical
potential \cite{Borsanyi:2014rza,Borsanyi:2013bia,Borsanyi:2013cga,Borsanyi:2010bp,Borsanyi:2010cj,Soltz:2015ula}. At non-zero quark
chemical potential $\mu_q \ne 0$, the primary quantities of interest  are the ``pressure
difference $\Delta P$'', the quark number density $n_B$ and quark
susceptibility  $\chi_q$ since these quantities are available from
lQCD  \cite{Borsanyi:2012cr,Endrodi:2011gv}. The lQCD results
can conveniently be interpreted within quasiparticle models
\cite{Gorenstein:1995vm,Levai:1997yx,Peshier:1999ww,Peshier:2002ww,Bannur:2005wm,Srivastava:2010xa,Bluhm:2004xn,Plumari:2011mk}
that are fitted to the equation of state from lQCD and
also allow for extrapolations to finite $\mu_q$, although with some
ambiguities. The quark number susceptibilities are additional
quantities to further quantify the properties of the partonic
degrees of freedom ( d.o.f.) especially in the vicinity of the QCD
phase transition or crossover
\cite{Bazavov:2009zn,Borsanyi:2010bp,Borsanyi:2010cj}.

Some early attempts to describe the lQCD pressure were based on the
notion of the QGP as a free gas of massless quarks and gluons
\cite{Fraga:2013qra} (Stephan-Boltzmann limit), or on the assumption
of interacting massless quarks and gluons following perturbative QCD
interactions \cite{Kajantie:2002wa}, or even as perturbative thermal
massive light quarks and gluons in the  Hard-Thermal-Loop (HTL)
approximation \cite{Peshier:2000hx}. These attempts failed to
reproduce lQCD results especially in the region 1 -- 3 $T_c$. Some
phenomenological models, based on the notion of the QGP as weakly
interacting quasi-particels (QPM) have been constructed to reproduce
the pressure and entropy from lQCD \cite{Bluhm:2004xn,Ratti:2007jf}.
Nevertheless, the challenge of describing simultaneously both the
lQCD pressure and quark susceptibilities as well as transport
coefficients is out of reach in these models \cite{Bluhm:2004xn},
especially if the quasi-particle model is not fitted to quark
susceptibilities but to the entropy density as common to most
approaches. Such findings have been pointed out before in Ref.
\cite{Plumari:2011mk} where the QPM underestimates the data on
susceptibilities since lattice results already reach the ideal gas
limit for temperatures slightly above $T_c$, leaving little space
for thermal parton masses. The apparent inconsistency between the
description of QCD thermodynamics and susceptibilities within the
standard quasi-particle model has been pointed out in particular in
Refs. \cite{Bluhm:2004xn,Plumari:2011mk}. Especially the quark
susceptibilities are very sensitive to the quark masses that are
used as inputs and solely determined by the quark degrees of
freedom. On the other hand both light quark and gluon masses
contribute to thermodynamic quantities like the entropy density and
pressure. Therefore, reconciling all observables from lQCD within a
single effective model is a challenge.

In this study we will consider the QGP as a dynamical quasi-particle
medium of massive off-shell particles (as described by the dynamical
quasiparticle model ``DQPM''
\cite{Cassing:2008nn,Bratkovskaya2011162,Peshier:2004ya}) and extend
the DQPM to partonic propagators that explicitly depend on the
three-momentum with respect to the partonic matter at rest in order
to match perturbative QCD (pQCD) at high momenta. We show that
within the extended DQPM -- denoted by DQPM$^*$ -- we reproduce the
lQCD equation of state at finite temperature $T$ and chemical
potential $\mu_q$. Moreover, we simultaneously describe the quark
number density and susceptibility $\chi_q$ from lQCD.
In the same approach, we also compute  the shear viscosity ($\eta$)
and electric conductivity ($\sigma_e$) of the QGP at finite
temperature and chemical potential in order to probe some transport
properties of the partonic medium in analogy to the earlier studies
in Refs.
\cite{Ozvenchuk:2012kh,Berrehrah:2013mua,Berrehrah:2014ysa,Berrehrah:2015ywa}.
The partonic spectral functions (or imaginary parts of the retarded
propagators) at finite temperature and chemical potential are
determined for these dynamical quasi-particles and the shear
viscosity $\eta$ is computed within the relaxation-time
approximation (RTA) which provides similar results as the Green-Kubo
method employed in Refs.
\cite{Peshier:2005pp,Aarts:2004sd,Aarts:2002cc}.

The paper is organized as follows:  We first present in Section
\ref{DQP-finiteTmu} the basic ingredients of the QGP d.o.f in terms
of their masses and widths which are the essential ingredients in
their retarded propagators as well as the running coupling (squared)
$g^2(T,\mu_q,p)$.  The gluon and fermion propagators -- as given by
the DQPM at finite momentum $p$, temperature $T$ and quark chemical
potential $\mu_q$ -- contain a few parameters that have to be fixed
in comparison to results from lQCD. The analysis of the lQCD
pressure and interaction measure in a partonic medium at finite $T$
and $\mu_q$ is performed in Section \ref{QGPthermo} while in Section
\ref{qChi} we investigate the quark number density and
susceptibility within the DQPM$^*$ and compare to lQCD results for
2+1 flavors ($N_f = 3$). In Section V we compute the QGP shear
viscosity and compare to lQCD results and other theoretical studies
while in Section VI we evaluate the electric conductivity of the
QGP. Throughout Sections \ref{QGPthermo}-\ref{etaOvs} we will point
out the importance of finite masses and widths of the dynamical
quasiparticles, including their finite momentum, temperature and
$\mu_q$ dependencies. In Section \ref{summary} we summarize the main
results and point out the future applications of the DQPM$^*$.

\section{Parton properties in the DQPM$^*$}
\label{DQP-finiteTmu}
In the DQPM$^*$ the entropy density $s(T)$, the pressure $P(T)$ and
energy density $\epsilon(T)$ are calculated in a straight forward
manner by starting with the entropy density in the quasiparticle
limit from Baym
\cite{Cassing:2008nn,Cassing:2008sv,Vanderheyden:1998ph},{\setlength\arraycolsep{0pt}
\begin{eqnarray}
\label{sdqp}
& & s^{dqp} = - d_g \!\int\!\!\frac{d \omega}{2 \pi} \frac{d^3p}{(2 \pi)^3} \frac{\partial n_B}{\partial T} \left( \Im\ln(-\Delta^{-1}) + \Im\Pi\,\Re\Delta \right)
\nonumber\\
& & {}    - d_q \!\int\!\!\frac{d \omega}{2 \pi} \frac{d^3p}{(2 \pi)^3} \frac{\partial n_F((\omega-\mu_q)/T)}{\partial T} \left( \Im\ln(-S_q^{-1}) + \Im \Sigma_q\ \Re S_q \right)
 \!
\nonumber\\
& & {}    - d_{\bar q} \!\int\!\!\frac{d \omega}{2 \pi} \frac{d^3p}{(2 \pi)^3} \frac{\partial n_F((\omega+\mu_q)/T)}{\partial T} \left( \Im\ln(-S_{\bar q}^{-1}) + \Im \Sigma_{\bar q}\ \Re S_{\bar q} \right) \!,
\end{eqnarray}} where $n_B(\omega/T) = (\exp(\omega/T)-1)^{-1}$ and
$n_F((\omega-\mu_q)/T) = (\exp((\omega-\mu_q)/T)+1)^{-1}$ denote the
Bose and Fermi distribution functions, respectively, while $\Delta
=(P^2-\Pi)^{-1}$, $S_q = (P^2-\Sigma_q)^{-1}$ and $S_{\bar q} =
(P^2-\Sigma_{\bar q})^{-1}$ stand for the full (scalar)
quasiparticle propagators of gluons $g$, quarks $q$ and antiquarks
${\bar q}$.  In Eq. (\ref{sdqp}) $\Pi$ and $\Sigma = \Sigma_q
\approx \Sigma_{\bar q}$ denote the (retarded) quasiparticle
selfenergies. In principle, $\Pi$ as well as $\Delta$ are Lorentz
tensors and should be evaluated in a nonperturbative framework. The
DQPM treats these degrees of freedom as independent scalar fields
with scalar selfenergies  which are assumed to be identical for
quarks and antiquarks. Note that one has to treat quarks and
antiquarks separately in Eq. (\ref{sdqp}) as their abundance differs
at finite quark chemical potential $\mu_q$. In Eq. (\ref{sdqp}) the
degeneracy for gluons is $d_g=2(N_c^2-1)$=16 while $d_q=d_{\bar
q}$=2$N_c N_f$=18 is the degeneracy for quarks and antiquarks with
three flavors. As a next step one writes the complex selfenergies as
$\Pi({\bsq q}) = M_g^2({\bsq q})-2i \omega \gamma_g({\bsq q})$ and
$\Sigma_{q} ({\bsq q}) = M_{q} ({\bsq q})^2 - 2 i \omega \gamma_{q}
({\bsq q})$ with a mass (squared) term $M^2$ and an interaction width
$\gamma$, i.e. the inverse retarded propagators ($\Delta, S_q$)
read,
\begin{equation}
\label{propdqpm} G_R^{-1} = \omega^2 - {\bsq q}^2 - M^2({\bsq q}) + 2i
\gamma({\bsq q}) \omega
\end{equation} and are analytic in the upper half plane in the
energy $\omega$. The imaginary part of (\ref{propdqpm}) then gives
the spectral function of the degree of freedom (except for a factor
$1/\pi$). In the DQPM
\cite{Cassing:2008nn,Bratkovskaya2011162,Peshier:2004ya} the masses
have been fixed in the spirit of the hard thermal loop (HTL)
approach with the masses being proportional to an effective coupling
$g(T/T_c)$ which has been enhanced in the infrared. In  the DQPM$^*$
the selfenergies depend additionally on the three-momentum $({\bs
q})$ with respect to the medium at rest, while the dependence on the
temperature $T/T_c$ and chemical potential $\mu_q$ are very similar
to the standard DQPM.

\subsection{Masses, widths and spectral functions of partonic
degrees of freedom in DQPM$^*$}

The functional forms for the partons masses and widths at finite
temperature $T$, quark chemical potential $\mu_q$
and momentum $p$ are assumed to be given by {\setlength\arraycolsep{0pt}
\begin{eqnarray}
\label{equ:Sec2.1} & & M_g (T, \mu_q, p) = \left(\frac{3}{2}\right)
\times \Biggl[\frac{g^2 (T^{\star}/T_c(\mu_q))}{6} \Bigl[\bigl( N_c
+ \frac{1}{2} N_f \bigr) T^2  + \frac{N_c}{2} \sum_q
\frac{\mu_q^2}{\pi^2} \Bigr] \times \Bigl[\frac{1}{1 + \Lambda_g
(T_c(\mu_q)/T^{\star}) p^2} \Bigr] \Biggr]^{1/2} + m_{\chi g}\; ,
\nonumber\\
& & {} M_{q,\bar{q}} (T, \mu_q, p)  = \Biggl[\frac{N_c^2-1}{8N_c}\,
g^2 (T^{\star}/T_c(\mu_q)) \Bigl[T^2 +  \frac{\mu_q^2}{\pi^2} \Bigr]
\times \Bigl[\frac{1}{1 + \Lambda_q (T_c(\mu_q)/T^{\star}) p^2}
\Bigr] \Biggr]^{1/2} + m_{\chi q} \, ,
\nonumber\\
& & {} \gamma_g (T, \mu_q, p) = N_c\, \frac{g^2
(T^{\star}/T_c(\mu_q))}{8\pi} \ T \, \ln \left(\frac{2 c}{g^2
(T^{\star}/T_c(\mu_q))}+1.1 \right)^{3/4} \times \Bigl[\frac{1}{1 +
\Lambda_g (T_c(\mu_q)/T^{\star}) p^2} \Bigr]^{1/2}\; ,
\nonumber\\
& & {} \gamma_{q,\bar{q}} (T, \mu_q, p) = \frac{N_c^2-1}{2N_c}\,
\frac{g^2 (T^{\star}/T_c(\mu_q))}{8\pi}\, T \ln \left(\frac{2 c}{g^2
(T^{\star}/T_c(\mu_q))}+1.1 \right)^{3/4} \times \Bigl[\frac{1}{1 +
\Lambda_q (T_c(\mu_q)/T^{\star}) p^2} \Bigr]^{1/2} \, ,
\end{eqnarray}}
where $T^{\star 2} = T^2 +  \mu_q^2/\pi^2$ is the effective
temperature used to extend the DQPM to finite $\mu_q$, $\Lambda_g
(T_c(\mu_q)/T^{\star})$ = 5  \ $(T_c(\mu_q)/T^{\star})^2$ GeV$^{-2}$
and $\Lambda_q (T_c(\mu_q)/T^{\star}) =$ 12 \
$(T_c(\mu_q)/T^{\star})^2$ GeV$^{-2}$. Here $m_{\chi g} \approx 0.5$
GeV is the gluon condensate and $m_{\chi q}$ is the light quark
chiral mass ($m_{\chi q} = 0.003$ GeV for $u$, $d$ quarks and
$m_{\chi q} = 0.06$ GeV for $s$ quarks). In Eq. (\ref{equ:Sec2.1})
$m_{\chi g}$ ($m_{\chi q}$) gives the finite gluon (light quark)
mass in the limit $p \rightarrow 0$ and $T = 0$ or for $p
\rightarrow \infty$. As mentioned above the quasiparticle masses and
widths (\ref{equ:Sec2.1}) are parametrized following hard thermal
loop (HTL) functional dependencies at finite temperature as in the
default DQPM \cite{Cassing:2008nn} in order to follow the correct
high temperature limit. The essentially new elements in
(\ref{equ:Sec2.1}) are the multiplicative factors specifying the
momentum dependence of the masses and widths with additional
parameters $\Lambda_g$ and $\Lambda_q$ and the additive terms
$m_{\chi g}$ and $m_{\chi q}$. The momentum-dependent factor in the
masses (\ref{equ:Sec2.1}) is motivated by Dyson-Schwinger studies in
the vacuum \cite{Fischer:2006ub} and yields the limit of pQCD for $p
\rightarrow \infty$.

The effective gluon and quark masses are a function of $T^{\star}$
at finite $\mu_q$. Here we consider three light flavors $(q = u, d,
s)$ and assume all chemical potentials to be equal $(\mu_u = \mu_d = \mu_s
= \mu_q)$. Note that alternative settings are also possible to
comply with strangeness neutrality in heavy-ion collisions. The coupling (squared) $g^2$ in (\ref{equ:Sec2.1})
is the effective running coupling given as a function of $T/T_c$ at $\mu_q = 0$. A straight
forward extension of the DQPM to finite $\mu_q$ is to consider the
coupling as a function of $T^{\star}/T_c (\mu_q)$ with a
$\mu_q$-dependent critical temperature $T_c (\mu_q)$,
{\setlength\arraycolsep{0pt}
\begin{eqnarray}
\label{equ:Sec2.2}
& & T_c (\mu_q) = T_c(\mu_q = 0) \sqrt{1 - \alpha \mu_q^2} \approx T_c(\mu_q = 0) \biggl(1 - \alpha/2 \mu_q^2 + \dots \biggr)
\end{eqnarray}}
with $\alpha \approx 8.79$ GeV$^{-2}$. We recall that the expression of
$T_c(\mu_q)$ in (\ref{equ:Sec2.2}) is obtained by requiring a
constant energy density $\epsilon$ for the system at $T =
T_c(\mu_q)$ where $\epsilon$ at $T_c(\mu_q = 0) \approx 0.158$ GeV
is fixed by lattice QCD calculation at $\mu_q = 0$. The coefficient
in front of the $\mu_q^2$-dependent part can
be compared to lQCD calculations at finite (but small) $\mu_B$ which
gives \cite{Bonati:2014rfa} {\setlength\arraycolsep{0pt}
\begin{eqnarray}
\label{equ:Sec2.3}
& & T_c (\mu_B) = T_c(\mu_B = 0) \biggl(1 - \kappa \left(\frac{\mu_B}{T_c (\mu_B = 0)}\right)^2 + \dots \biggr)
\end{eqnarray}}
with $\kappa = 0.013(2)$. Rewriting (\ref{equ:Sec2.2}) in the form
(\ref{equ:Sec2.3}) and using $\mu_B \approx 3 \mu_q$ we get
$\kappa_{DQPM} \approx 0.0122$ which compares very well with the
lQCD result.

Using the pole masses and widths (\ref{equ:Sec2.1}), the spectral
functions for the partonic degrees of freedom are fully determined,
i.e. the imaginary parts of the retarded propagators. The real part
of the retarded propagators then follows from dispersion relations.
Since the retarded propagators show no poles in the upper complex
half plane in the energy $\omega$ the model propagators obey
micro-causality \cite{Rauber:2014mca}. The imaginary parts are of
 Lorentzian form and provide the spectral functions $\rho_i (p)$ with $p =
(\omega, \bsp{p})$
\cite{Cassing:2008nn,Cassing:2009vt,Cassing2007365},
\begin{equation}
\label{equ:Sec2.4} \rho_i (\omega, \bsp{p} ) = \frac{4 \omega
\gamma_i({\bs p})}{(\omega^2 - \bs{p}^2 - M_i^2({\bs p}))^2 + 4
\gamma_i^2({\bs p}) \omega^2}
 \equiv \frac{\gamma_i({\bs p})}{\tilde{E}_i({\bs p})} \biggl(\frac{1}{(\omega - \tilde{E}_i({\bs p}))^2 + \gamma_i^2({\bs p})} -
\frac{1}{(\omega + \tilde{E}_i({\bs p}))^2 + \gamma_i^2({\bs p})} \biggr)
\end{equation}
with $ \tilde{E}_i^2 (\bs{p}) = \bs{p}^2 + M_i^2({\bs p}) -
\gamma_i^2({\bs p})$ for $i \in [g, q, \bar{q}]$. These spectral
functions (\ref{equ:Sec2.4}) are antisymmetric in $\omega$ and
normalized as
{\setlength\arraycolsep{-10pt}
\begin{eqnarray}
\label{equ:Sec2.5} & & \hspace{-0.6cm} \int_{- \infty}^{+ \infty}
\frac{d \omega}{2 \pi} \ \omega \ \rho_i (\omega, \bs{p}) =
\int_0^{+ \infty} \frac{d \omega}{2 \pi} \ 2 \omega \ \rho_i
(\omega, \bs{p}) = 1.
\end{eqnarray}}
where $M_i(T,\mu_q,{\bs p})$, $\gamma_i(T,\mu_q,{\bs p})$ are the
particle pole mass and width at finite three momentum ${\bs p}$,
temperature $T$ and chemical potential $\mu_q$, respectively.

\subsection{The running coupling in DQPM$^*$}
In contrast to our previous DQPM studies in Refs.
\cite{Berrehrah:2013mua,Berrehrah:2014ysa,Berrehrah:2015ywa} we
suggest here a new solution for the determination of the effective
coupling  which is more flexible. Our new strategy to determine
$g^2(T(T_c)$ is the following: For every temperature $T$ we fit the
DQPM$^*$ entropy density (\ref{sdqp}) to the entropy density
$s^{lQCD}$ obtained by lQCD. In practice, it has been checked that
for a given value of $g^2$, the ratio $s(T,g^2)/T^3$ is almost
constant for different temperatures and identical to $g^2$. Moreover
$\frac{\partial}{\partial T} (s(T,g^2)/T^3) = 0$. Therefore the
entropy density $s$ and the dimensionless equation of state in the
DQPM$^*$ is a function of the effective coupling only, i.e.
$s(T,g^2)/s_{SB} (T) = f(g^2)$. The functional form  $$\displaystyle
f(g^2) = \frac{1}{(1 + a_1 . (g^2)^{a_2})^{a_3}}$$ is suited to
describe $s^{lQCD}(T,g^2)/s_{SB}$. By inverting $f(g^2)$, one
arrives at the following parametrization for $g^2$ as a function of
$s/s_{SB}$: {\setlength\arraycolsep{0pt}
\begin{eqnarray}
\label{equ:Sec2.6} g^2(s/s_{SB}, T)  \sim  \left( \frac{a}{T} + b
\right)  \Biggl(\left(\frac{s/s_{SB}}{d (T)} \right)^{v(T)} - 1
\Biggr)^{w(T)},
\end{eqnarray}}
with  $S_{SB}^{QCD} = 19/(9\pi^2 T^3)$. Since the entropy density
from lQCD has the proper high temperature limit, the effective
coupling $g^2$ also gives the correct asymptotics for $T \rightarrow
\infty$ and decreases as $g^2 \sim 1/\log(T^2)$. The
temperature-dependent parameters $v(T)$, $w(T)$ and $d (T)$ all have
the functional form: {\setlength\arraycolsep{0pt}
\begin{eqnarray}
\label{equ:Sec2.61}
f (T)  =  \frac{a}{(T^b + c)^d} . (T + e),
\end{eqnarray}}
where the parameters $a$, $b$, $c$, $d$ and $e$ are fixed once for
each function $v(T)$, $w(T)$ and $d (T)$.

Note that with the parametrization (\ref{equ:Sec2.6}) for
$g^2(s/s_{SB}, T)$ one can easily adapt to any equation of state and
therefore avoid a refitting of the coupling in case of new (or
improved) lattice data. However, the coupling (\ref{equ:Sec2.6}) is
valid only for a given number of quark flavors $N_f$ which is fixed
by the lQCD equation of state.

To obtain $g^2(T/T_c)$ from $g^2(s/s_{SB}, T)$, we proceed as
follows:
\begin{itemize}
\item Using the equation of state from the Wuppertal-Budapest collaboration
\cite{Borsanyi:2012cr},
which provide an analytical parametrization of the interaction
measure $I/T^4$, {\setlength\arraycolsep{0pt}
\begin{eqnarray}
\label{equ:Sec2.7}
\frac{I(T)}{T^4} = \exp(-h_1/t - h_2/t^2). \Biggl(h_0 + \frac{f_0 (\tanh(f_1 . t + f_2) + 1)}{1 + g_1 . t + g_2 . t^2} \Biggr),
\end{eqnarray}}
 with $t = T/200$ MeV, $h_0 = 0.1396$, $h_1 = - 0.18$, $h_2 = 0.035$, $f_0 = 2.76$, $f_1 = 6.79$, $f_2 = - 5.29$, $g_1 = - 0.47$ and $g_2 = 1.04$,
\item we calculate the pressure $P/T^4$ by
{\setlength\arraycolsep{0pt}
\begin{eqnarray}
\label{equ:Sec2.8}
\frac{P(T)}{T^4} = \int_0^T \frac{I(T_0)}{T_0^5} d T_0,
\end{eqnarray}}
\item and then the entropy density $s/s_{SB}$
{\setlength\arraycolsep{0pt}
\begin{eqnarray}
\label{equ:Sec2.9}
s/s_{SB} = \frac{I(T)/T^4 + 4 P/T^4}{19/(9\pi^2)}.
\end{eqnarray}}
\item Replacing $s/s_{SB}$ from Eq.(\ref{equ:Sec2.9}) in  Eq.(\ref{equ:Sec2.6}) we obtain $g^2(T/T_c)$.
\end{itemize}
The procedure outlined above yields $g^2(T/T_c)$ for $\mu_q=0$. For finite
$\mu_q$ we will make use of $g^2 (T/T_c) \rightarrow g^2
(T^{\star}/T_c(\mu_q))$, with the $\mu_q$-dependent critical
temperature $T_c (\mu_q)$ taken from (\ref{equ:Sec2.2}). The running
coupling (\ref{equ:Sec2.6})-(\ref{equ:Sec2.9}) permits for an
enhancement near $T_c$ as already introduced in Ref. \cite{Peshier:2002ww}.

\begin{figure}[h!] 
\begin{center}
\begin{minipage}{13.8pc}
\includegraphics[width=13.8pc, height=14.5pc]{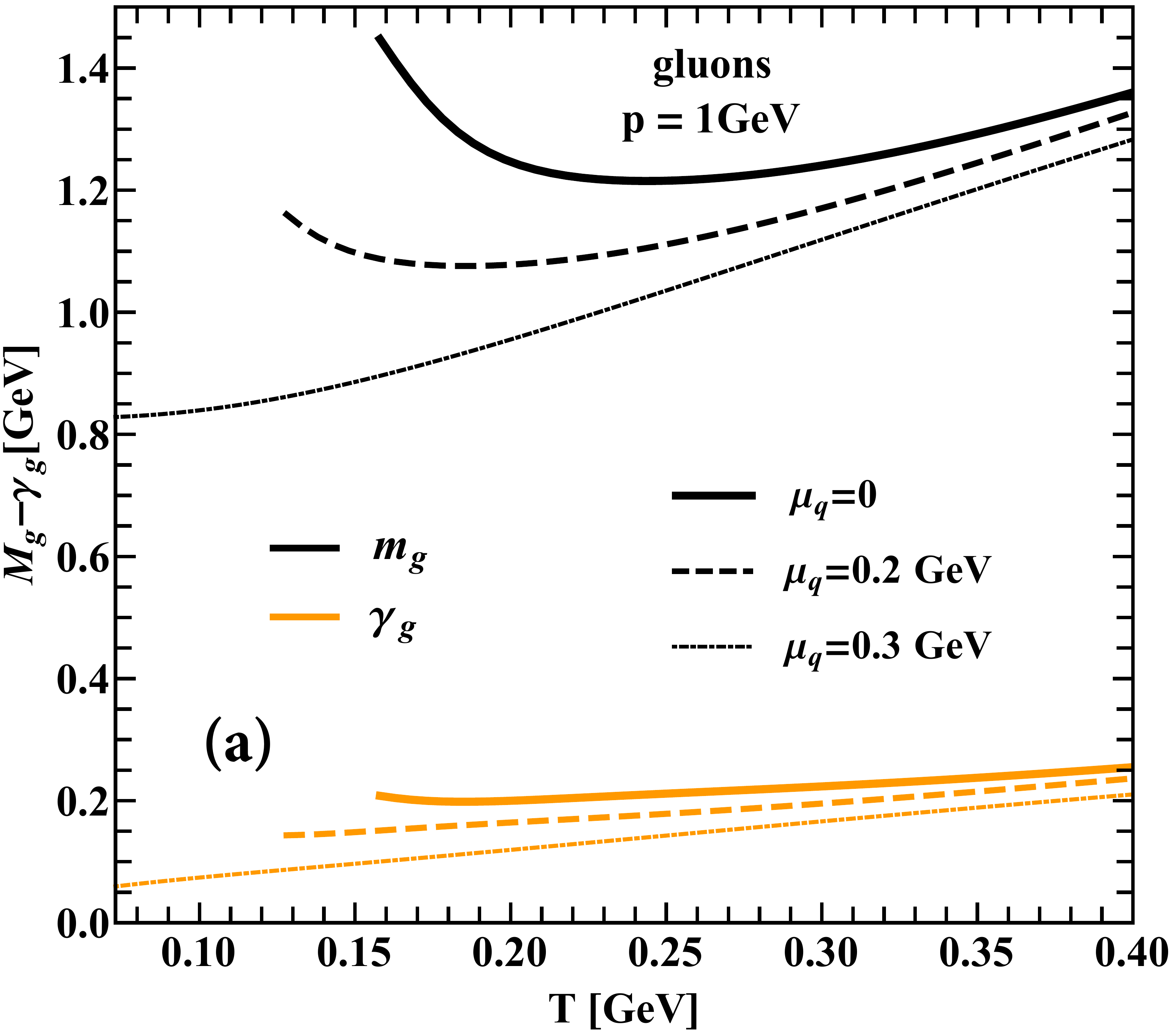}
\end{minipage} \hspace*{0.03cm}
\begin{minipage}{13.8pc}
\includegraphics[width=13.8pc, height=14.5pc]{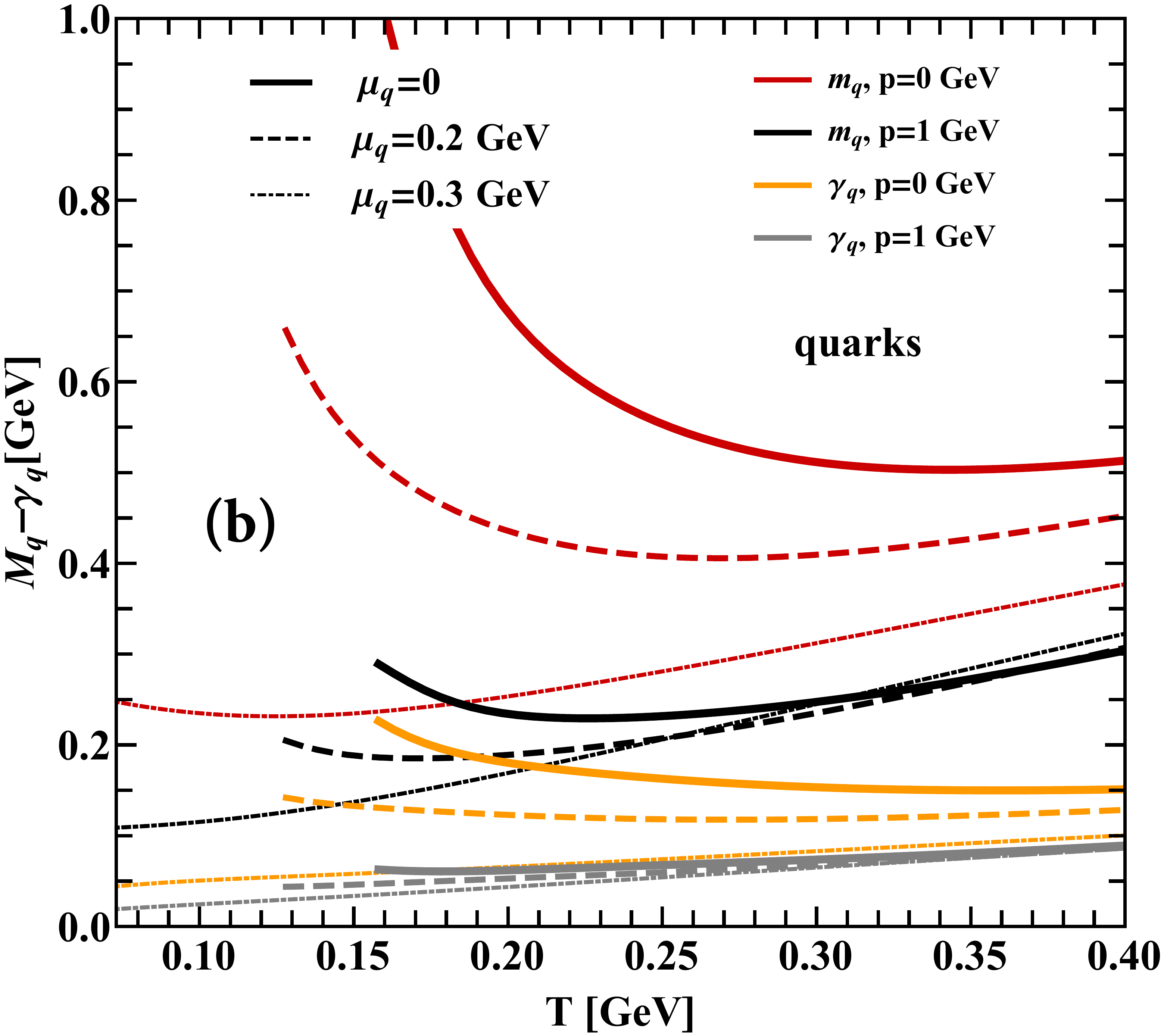}
\end{minipage} \hspace*{0.03cm}
\begin{minipage}{13.8pc}
\includegraphics[width=13.8pc, height=14.5pc]{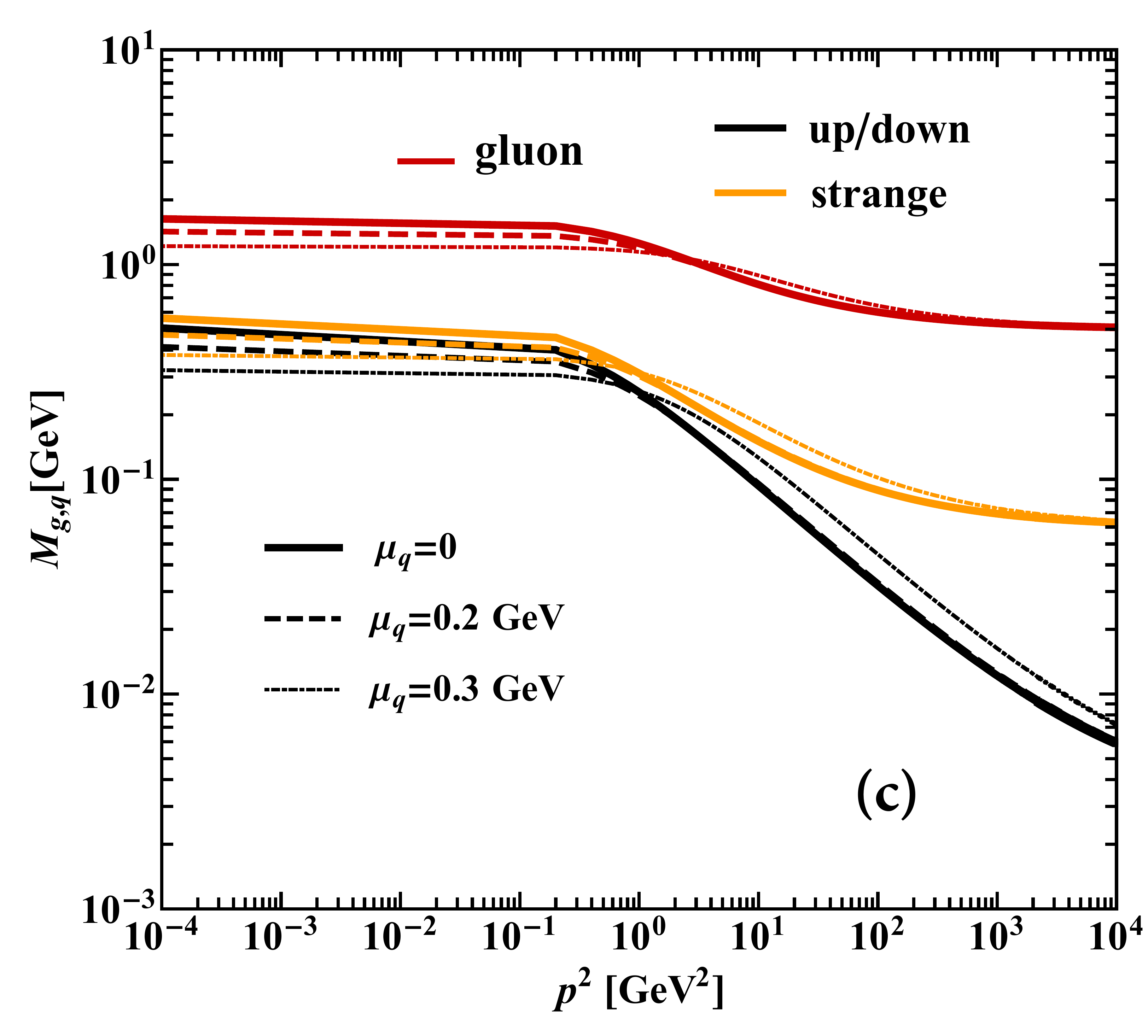}
\end{minipage}
\end{center}
\vspace*{-0.5cm}
\caption{\emph{(Color online) The DQPM$^*$ gluon (a) and light quark (b) masses and widths given by (\ref{equ:Sec2.1})
using the coupling  (\ref{equ:Sec2.6})-(\ref{equ:Sec2.9}) for different quark chemical potentials as a
function of the temperature $T$. (c) Gluon and light quark masses as a function of the momentum squared for $T = 2 T_c$
and $\mu_q = 0, 0.2, 0.3$ GeV.}}
\label{fig:mpDQPMIngred}
\end{figure}

Figs. \ref{fig:mpDQPMIngred}, (a)-(b) show the gluon and light quark
masses and widths, respectively, at finite temperature and chemical
potential for a momentum $p = 1$ GeV/c. Furthermore, Fig.
\ref{fig:mpDQPMIngred} (c) shows the gluon and light quark masses as
a function of momentum (squared) $p^2$ at finite temperature $T = 2
T_c$ and different $\mu_q$. Note that for $p=0$ we obtain  higher
values of the gluon and light quark masses (as a function of $T$ and
$\mu_q$) since for finite  momenta the masses decrease (at a given
temperature and chemical potential), especially for the light quarks
as seen in Fig \ref{fig:mpDQPMIngred} (c). The extension $T/T_c
\rightarrow T^{\star}/T_c(\mu_q)$ for finite $\mu_q$ in the
functional form for the strong coupling leads to lower values for
the parton masses and widths at finite $\mu_q$ as compared to
$\mu_q=0$ near $T_c (\mu_q)$.

\section{Thermodynamics of the QGP from DQPM$^*$}
\label{QGPthermo}

The expressions for the equation of state (energy density
$\epsilon$, entropy density $s$ and pressure $P$) of strongly
interacting matter have been given for finite temperature and
chemical potential in Ref. \cite{Marty:2013ita} for on-shell partons
and in \cite{Cassing:2008nn} for the case of off-shell partons using
the relations based on the stress-energy tensor $T^{\mu \nu}$. We
recall that the approach for calculating the equation of state in
the DQPM$^*$ is based on thermodynamic relations (see below). The
procedure is as follows: One starts from the evaluation of the
entropy density $s$ from (\ref{sdqp}) employing the masses and
widths obtained from the expressions (II.1). Then using the
thermodynamic relation $s = ({\partial P}/{\partial T})_{\mu_q}$
(for a fixed quark chemical potential $\mu_q$) one obtains the
pressure $P$ by integration of $s$ over $T$ while the energy density
$\epsilon$ can be gained using the relation,
\begin{equation} T s (T, \mu_B) = \epsilon (T, \mu_B) + P (T, \mu_B) - \mu_B
n_B (T, \mu_B) , \end{equation} where $n_B$ is the net baryon
density.

\begin{figure}[h!] 
\begin{center}
\begin{minipage}{18pc}
\includegraphics[width=18pc, height=17pc]{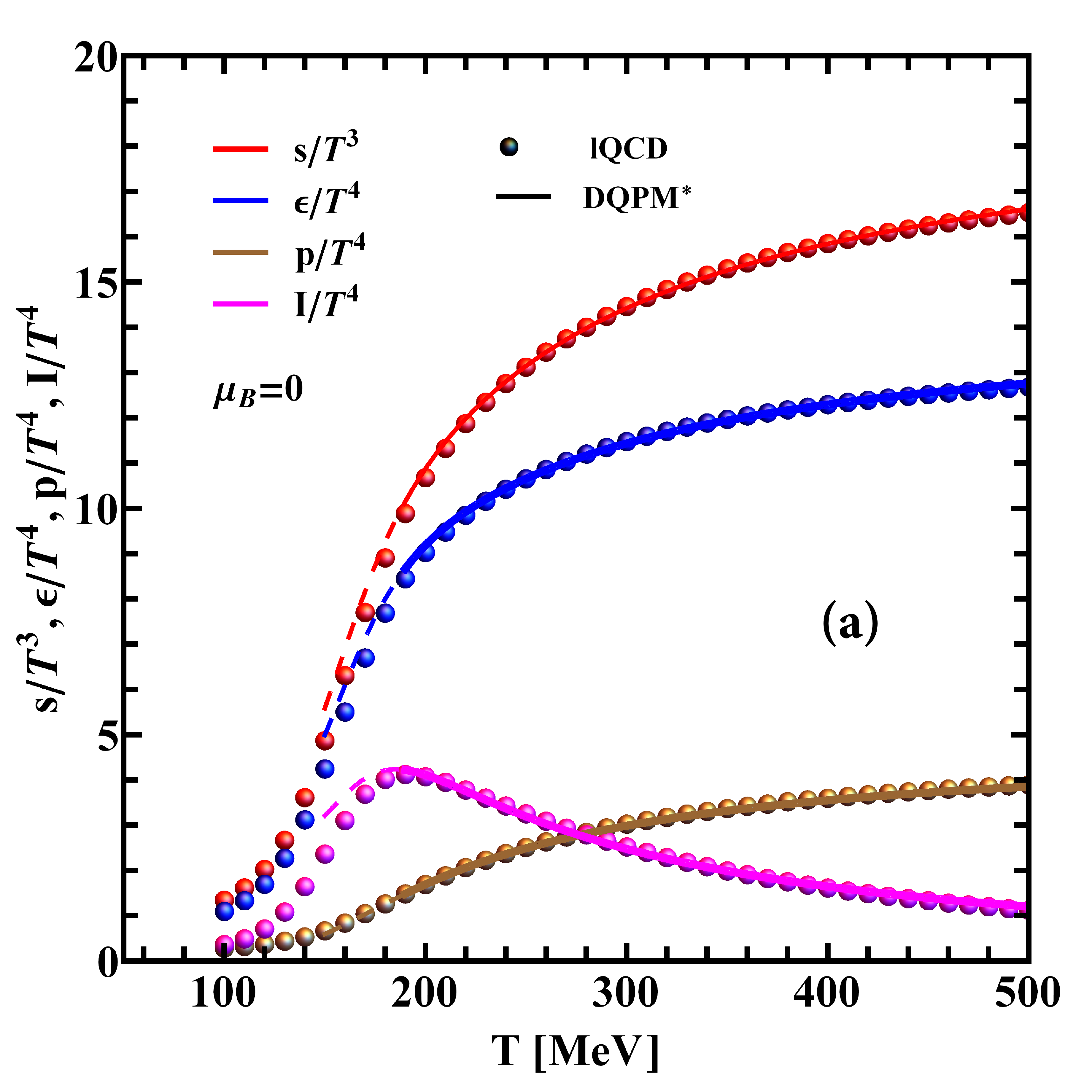}
\end{minipage}\hspace*{0.1cm}
\begin{minipage}{19pc}
\includegraphics[width=18pc, height=17pc]{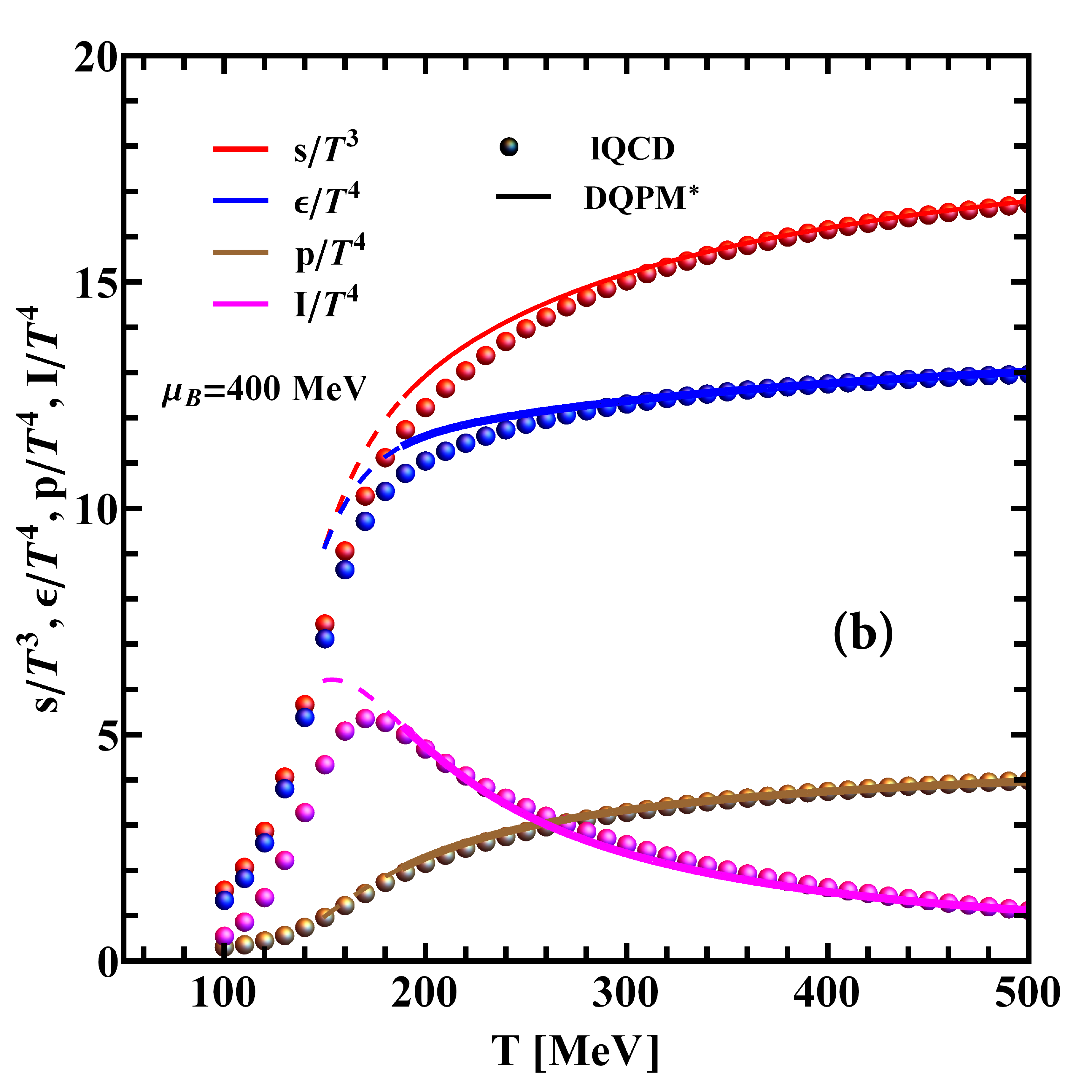}
\end{minipage}
\end{center}
\caption{\emph{(Color online) Energy density $\epsilon$, entropy
density $s$, pressure density $P$ and trace anomaly $(I = \epsilon -
3 P)$ as a function of temperature $T$ at $\mu_B = 0$ (a) and at
$\mu_B=400$ MeV (b) from DQPM$^*$ compared to lQCD data from Ref.
\cite{Borsanyi:2012cr}.}} \label{fig:EoSTmu}
\end{figure}

The energy density $\epsilon$, entropy density $s$, pressure $P$ and
the interaction measure $[ I (T, \mu_q) = \epsilon (T, \mu_q) - 3 P
(T, \mu_q) ]$ --known in lQCD as the trace anomaly-- in the DQPM$^*$
are shown in Fig.\ref{fig:EoSTmu} (a), (b) as a function of
temperature $T$  for two values of the baryon chemical potential
$\mu_B = 0$ and $\mu_B = 400$ MeV, respectively (where $\mu_B = 3
\mu_q$ in our study). We, furthermore, compare our results with
lattice calculations from Ref. \cite{Borsanyi:2012cr}. We notice
that our results are in a very good agreement with the lattice data
for $\mu_B = 0$ (a) and in case of $\mu_B = $ 400 MeV (b) for
temperatures larger than $1.2 \ T_c (\mu_q)$. In the latter case we
observe (for temperatures just above $T_c (\mu)$) some deviations
which are expected to result from additional hadronic degrees of
freedom in the crossover region.  The excess in quarks can be seen
also in the net baryon density $n_B$, as we will show below.

At finite baryon chemical potential i.e. $\mu_B = 400$ MeV, the
maximum of the trace anomaly  is shifted towards  lower
temperatures. We notice also the proper scaling of our DQPM$^*$
description of QGP thermodynamics, when moving from zero to finite
quark chemical potential (cf. Fig.\ref{fig:EoSTmu} (a) and (b)).

\section{Quark number density and susceptibility from DQPM$^*$}
\label{qChi}

\subsection{Baryon number density in the DQPM$^*$} 

The equation of state for vanishing chemical potential is defined
solely by the entropy density; for finite chemical potential one has
to include the particle density. In the DQPM$^*$ the quark density
$n^{dqp}$ in the quasiparticle limit is defined in analogy to the
entropy density (\ref{sdqp}) as
\cite{Blaizot:2000fc},{\setlength\arraycolsep{0pt}
    \begin{eqnarray}
    \label{ndqp}
    n^{dqp} = & & \ - d_q \!\int\!\!\frac{d \omega}{2 \pi} \frac{d^3p}{(2 \pi)^3} \frac{\partial n_F((\omega-\mu_q)/T)}{\partial \mu_q} \left( \Im\ln(-S_q^{-1}) + \Im \Sigma_q\ \Re S_q \right)
    \!
    \nonumber\\
    & & {}    - d_{\bar q} \!\int\!\!\frac{d \omega}{2 \pi} \frac{d^3p}{(2 \pi)^3} \frac{\partial n_F((\omega+\mu_q)/T)}{\partial \mu_q} \left( \Im\ln(-S_{\bar q}^{-1}) + \Im \Sigma_{\bar q}\ \Re S_{\bar q} \right) \!,
    \end{eqnarray}}
and $n_B$ from (\ref{ndqp}) is split following the on-shell $n_B^{(0)}$  
and off-shell $\Delta n_B$ terms, with $n_B = n_B^{(0)}+ \Delta n_B$ as: {\setlength\arraycolsep{-1pt} 
\begin{eqnarray}
\label{equ:Sec4.3bis}
& &  n_B^{(0)} = d_q \int \frac{d^3 p}{(2 \pi)^3} \ f_q^{(0)} - d_{\bar{q}} \int \frac{d^3 p}{(2 \pi)^3} \ f_{\bar{q}}^{(0)},
\end{eqnarray}}
{\setlength\arraycolsep{0pt}
\begin{eqnarray}
\label{equ:Sec4.3tr}
& & \Delta n_B = \int \frac{d \omega}{(2 \pi)} \frac{d^3 p}{(2 \pi)^3} \frac{\partial f_q ((\omega - \mu_q)/T)}{\partial \mu_q} \Biggl(2 \gamma \omega \frac{\omega^2 - \textbf{p}^2 - M^2}{(\omega^2 - \textbf{p}^2 - M^2)^2 + 4 \gamma^2 \omega^2} - \arctan \left(\frac{2 \gamma \omega}{\omega^2 - \textbf{p}^2 - M^2} \right) \Biggr)
\nonumber\\
& & {} + \int \frac{d \omega}{(2 \pi)} \frac{d^3 p}{(2 \pi)^3} \frac{\partial f_{\bar{q}} ((\omega + \mu_q)/T)}{\partial \mu_q} \Biggl(2 \gamma \omega \frac{\omega^2 - \textbf{p}^2 - M^2}{(\omega^2 - \textbf{p}^2 - M^2)^2 + 4 \gamma^2 \omega^2} - \arctan \left(\frac{2 \gamma \omega}{\omega^2 - \textbf{p}^2 - M^2} \right) \Biggr),
\end{eqnarray}}
where $f_q^{(0)} = (\exp ((\sqrt{p^2 + M^2} - \mu_q)/T)+1)^{-1}$,
$f_{\bar{q}}^{(0)} = (\exp ((\sqrt{p^2 + M^2} + \mu_q)/T)+1)^{-1}$
denote again the Fermi distribution functions for the on-shell quark
and anti-quark, with $M$ corresponding to the pole mass. The on- and off-shell terms can be interpreted as arising from
a pole- and a damping-term, respectively. The pole term $n_B^{(0)}$
corresponds to the baryon density of a non-interacting massive gas
of quasiparticles, whereas the additional contribution due to the
damping term $\Delta n_B$ has to be attributed to the  finite width
of the quasiparticles.

Finally, note that the quark number density follows from the same
potential as the entropy density \cite{Vanderheyden:1998ph} which
ensures that it fulfills the thermodynamic relation $n=(\partial
P/\partial \mu_q)_T$ (for fixed temperature T). To be fully
thermodynamically consistent the entropy and the particle density
have to satisfy the Maxwell relation $(\partial n/\partial
T)_{\mu_q}=(\partial s/\partial \mu_q)_T$. This provides further
constraints on the effective coupling $g^2(T,\mu_q)$ at finite
chemical potential which we neglect in the current approach.
Nevertheless, it was checked that the violation of the Maxwell
relation is generally small and most pronounced around $T_c$. We
note, however, that when extending the approach to even larger
chemical potentials the full thermodynamic consistency has to be
taken into account. The baryon number density, finally, is related
to the quark number density by the simple relation $n_B=n^{dqp}/3$.

\subsection{Susceptibilities in the DQPM$^*$}

From the densities $n_B$ one may obtain other thermodynamic
quantities like the pressure difference $\Delta P$ and the quark
susceptibilities $\chi_q$, which can be confronted with lattice data
for $N_f=2$ from Alton \emph{et al}.
\cite{Allton:2003vx,Allton:2005gk} and for $N_f = 3$ from Borsanyi
\emph{et al}.\cite{Borsanyi:2012cr}. We recall that the quark-number
susceptibility measures the static response of the quark number
density to an infinitesimal variation of the quark chemical
potential. From (\ref{equ:Sec4.3bis})-(\ref{equ:Sec4.3tr}) we calculate
$\Delta P$ and $\chi_q$ as {\setlength\arraycolsep{0pt}
\begin{eqnarray}
\label{equ:Sec4.4}
& & \Delta P (T, \mu_B) \equiv  P (T, \mu_B) - P (T,0) = \int_0^{\mu_B} n_B \ d \mu_B.
\end{eqnarray}}
{\setlength\arraycolsep{0pt}
\begin{eqnarray}
\label{equ:Sec4.5} & & \chi_q (T) = \frac{\partial n_q}{\partial
\mu_q} \biggl|_{\mu_q=0}; \hspace*{0.5cm} \chi_q (T, \mu_q) =
\frac{1}{9} \frac{\partial n_B}{\partial \mu_B}.
\end{eqnarray}}
Furthermore, for small $\mu_q$ a Taylor expansion of the pressure in
$\mu_q/T$ can be performed which gives {\setlength\arraycolsep{0pt}
\begin{eqnarray}
\label{equ:Sec4.6}
& & \frac{P (T, \mu_q)}{T^4} = \sum_{n=0}^{\infty} c_n (T) \left(\frac{\mu_q}{T} \right)^n,
\hspace*{0.5cm}  c_n (T) = \frac{1}{n!} \frac{\partial^n (P(T, \mu_q)/T^4)}{\partial (\mu_q/T)^n}\Biggr|_{\mu_q = 0},
\end{eqnarray}}
where $c_n(T)$ is vanishing for odd $n$ and $c_0 (T)$ is given by
$c_0 (T) = p (T, \mu_q = 0)$. As shown above the  DQPM$^*$ compares
well with lattice QCD results for $c_0(T)$. Since $\chi_q$ at finite
$\mu_q$ is related to the pressure by $$\displaystyle \chi_q (T,
\mu_q)/T^2 =
\partial^2 (P/T^4)/\partial^2 (\mu_q/T) ,$$ one can define the
susceptibility $\chi_2^{i j}$ at vanishing quark chemical potential
as \cite{Borsanyi:2012cr} {\setlength\arraycolsep{0pt}
\begin{eqnarray}
\label{equ:Sec4.7}
& & \frac{P(T,{\mu_i})}{T^4} = \frac{P(T, {0})}{T^4} + \frac{1}{2} \sum_{i, j} \frac{\mu_i \mu_j}{T^2} \chi_2^{i j},
\hspace*{0.5cm} \textrm{with} \ \ \chi_2^{i j} = \frac{1}{T^2} \frac{\partial n_j (T, {\mu_i})}{\partial \mu_i} \Biggr|_{\mu_i = \mu_j = 0},
\end{eqnarray}}
which in case of 3 flavors ($u$, $d$, $s$ quarks) with $\mu_u =
\mu_d = \mu_s$ becomes {\setlength\arraycolsep{0pt}
\begin{eqnarray}
\label{equ:Sec4.8} & & \chi_2 (T) = \frac{1}{9} \ \frac{1}{T^2}
\frac{\partial n_q (T, \mu_q)}{\partial \mu_q} \Biggr|_{\mu_q = 0} =
\frac{1}{9} \ \frac{\chi_q (T)}{T^2}.
\end{eqnarray}}
We recall again that the susceptibilities are the central quantities
in lQCD calculations for nonzero $\mu_q$.

\subsection{$n_B$ and $\chi_q$: DQPM$^*$ \emph{vs} lQCD}

\begin{figure}[h!] 
\begin{center}
\begin{minipage}{18pc}
\includegraphics[width=18pc, height=17.5pc]{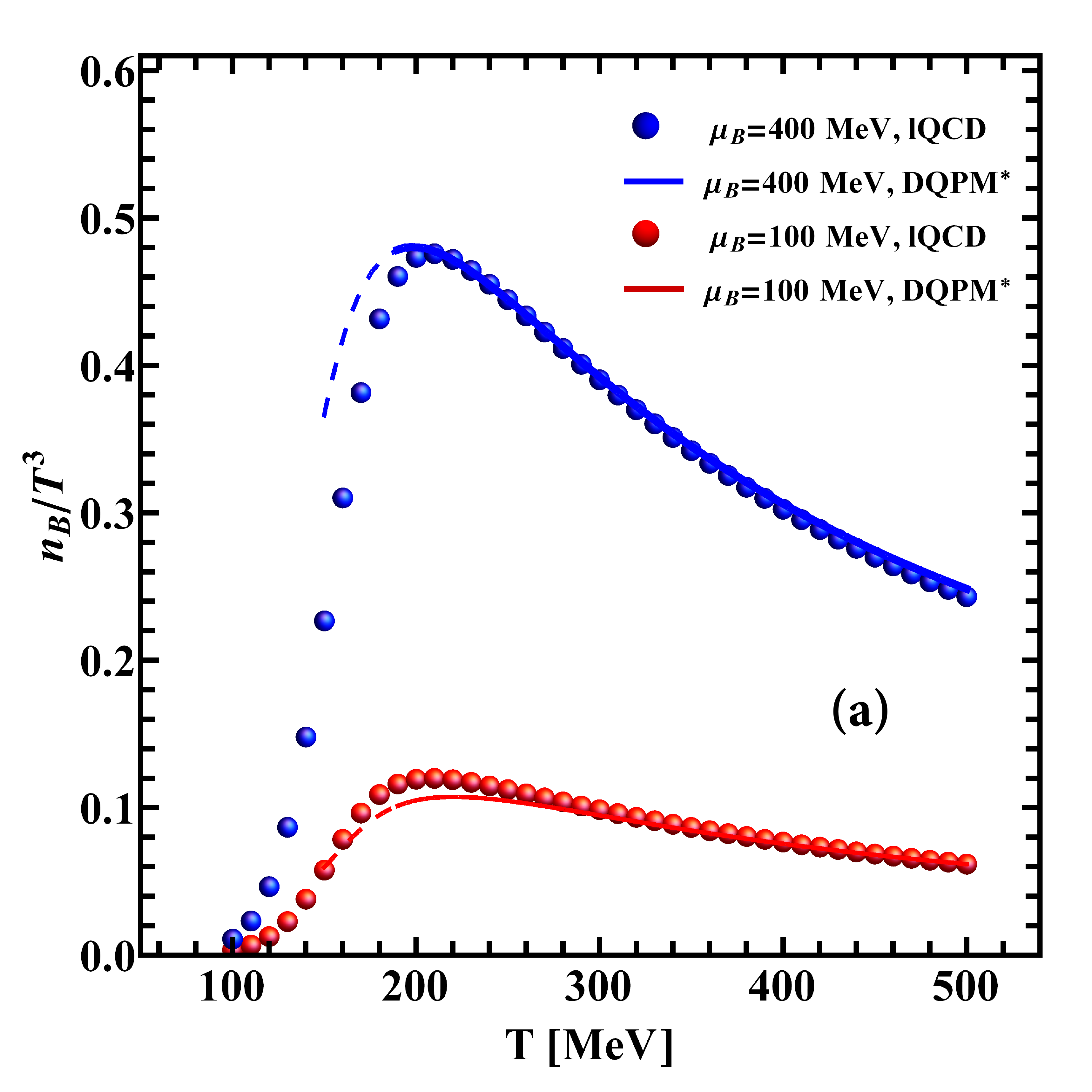}
\end{minipage}\hspace*{0.5cm}
\begin{minipage}{18pc}
\includegraphics[width=18pc, height=17.5pc]{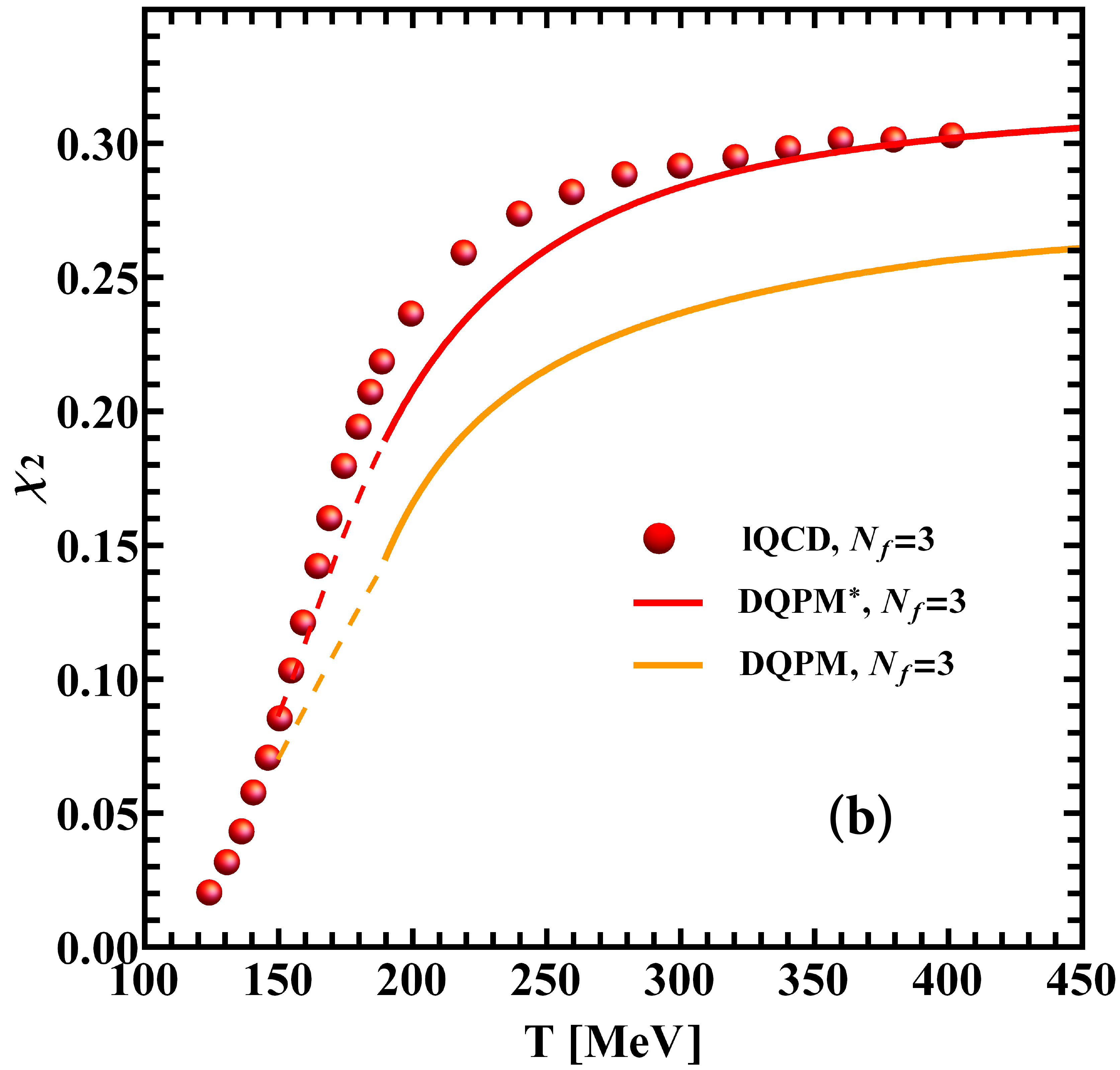}
\end{minipage}
\end{center}
\vspace*{-0.5cm} \caption{\emph{(Color online) (a) The baryon number
density $n_B$ from DQPM$^*$ as compared to lattice data from
\cite{Borsanyi:2012cr} for $N_f = 3$ for quark chemical potential
$\mu_q = 0$. (b) The susceptibility $\chi_2$ from DQPM$^*$ as
compared to lattice data from \cite{Borsanyi:2012cr} for $N_f = 3$
and $\mu_q=0$ using Eq. (\ref{equ:Sec4.8}). The lower (orange) line
gives the result from the conventional DQPM, i.e. with momentum
independent masses.}} \label{fig:nBchi2Nf3}
\end{figure}

Using the masses and widths (\ref{equ:Sec2.1}) and the running
coupling (\ref{equ:Sec2.6})-(\ref{equ:Sec2.8}), we calculate the
baryon number density $n_B$
(\ref{equ:Sec4.3bis})-(\ref{equ:Sec4.3tr}) and quark susceptibility
$\chi_2$ including the finite width of the parton spectral
functions. The results for $n_B$ and $\chi_2$ for $N_f = 3$ are
given in Fig.\ref{fig:nBchi2Nf3} (a) and (b), respectively. The
comparison with the lattice data from \cite{Borsanyi:2012cr} is
rather good which is essentially due to an extra contribution
arising from the momentum dependence of the DQPM$^*$ quasiparticles
masses and widths. Such a momentum dependence in $m_{q,\bar{q},g}$
and $\gamma_{q,\bar{q},g}$ decreases the 'thermal average' of light
quark and gluon masses which improves the description of lQCD
results for the susceptibilities. For comparison we also show the
result for $\chi_q$ from the conventional DQPM, i.e. with momentum
independent masses, which substantially underestimates the lattice
data. The small difference between lQCD and DQPM$^*$ for $n_B$ and
$\chi_2$ close to $T_c$ is related to a possible excess of light
quarks and antiquarks  which should combine to hadrons in the
crossover region. We recall that the DQPM$^*$ describes only the QGP
phase and deals with dynamical quarks and gluons solely.

Finally, we emphasize the challenge to describe simultaneously the
entropy $s$ and pressure $P$ on one side and $n_B$ and $\chi_2$ on
the other side. Indeed, increasing the light quark mass and width
helps to improve the description of $s$ and $P$ (for $\mu_B$ = 400
MeV), but this leads to a considerable decrease in $n_B$ and
$\chi_2$. In other words, lighter quarks are favorable to improve
the agreement with lQCD data on $n_B$ and $\chi_2$, however, this
leads to an increase of $s$ and $P$, which can be only partially
counterbalanced by an increasing gluon mass and width.

\section{Shear viscosity of the QGP from DQPM$^*$}
\label{etaOvs}

In this Section we focus on the shear viscosity of the QGP using the
relaxation time approximation (RTA). In the dilute gas approximation
the relaxation time $\tau_i$ of the particle $i$ is obtained for on-
or off-shell quasi-particles by means of the partonic scattering
cross sections, where the $qq$, $q\bar{q}$, $qg$ and $gg$ elastic
scattering processes as well as some inelastic processes involving
chemical equilibration, such as $gg \rightarrow q \bar{q}$ are
included in the computation of $\tau_i$ \cite{Berrehrah:2014ysa}.
For the DQPM$^*$ approach  we do not need the explicit cross
sections since the inherent quasi-particle width $\gamma_i (T,\mu_q,
p)$ directly provides the total interaction rate
\cite{Cassing:2008nn}. To this end we only have to evaluate the
average of the momentum dependent widths $\gamma_g (T,\mu_q, p)$ and
$\gamma_q (T,\mu_q, p)$ over the thermal distributions at fixed $T$
and $\mu_q$, i.e. $\bar \gamma_g (T,\mu_q)$ and $\bar \gamma_q
(T,\mu_q)$.

The shear viscosity $\eta (T,\mu_q)$ is defined in the dilute gas
approximation for the case of off-shell particles by
\cite{Chakraborty:2010fr,Berrehrah:2014ysa}
{\setlength\arraycolsep{-8pt}
\begin{eqnarray}
\label{equ:Sec5.1}
& & \hspace*{-0.7cm} \eta (T,\mu_q) = \frac{1}{15 T} d_g \!\!\int\!\! \frac{d^3 p}{(2\pi)^3} \ \!\!\int\!\!
\frac{d\omega}{2 \pi} \omega \ \bar \tau_g (T, \mu_q) \ f_g(\omega/T) \times \rho_{g}
(\omega, {\bs p}) \frac{{\bs p}^4}{\omega^2} \Theta(P^2)
\nonumber\\
& & {} \hspace*{-0.7cm} + \frac{1}{15 T} \frac{d_q}{6} \!\!\int\!\! \frac{d^3
p}{(2\pi)^3} \ \!\!\int\!\! \frac{d \omega}{2 \pi} \omega \  \Bigg[
\sum_q^{u,d,s} \bar \tau_q (T, \mu_q) f_q((\omega-\mu_q)/T)
\rho_{q}(\omega,{\bs p})+ \sum_{\bar q}^{\bar u,\bar d,\bar s}
\bar \tau_{\bar q} (T, \mu_q) f_{\bar q}((\omega+\mu_q)/T) \ \rho_{\bar
q}(\omega,{\bs p}) \  \Bigg] \frac{{\bs p}^4}{\omega^2} \Theta(P^2),
\end{eqnarray}}
where ${\bs p}$ is the three-momentum and $P^2$ the invariant mass
squared. The functions $\rho_g, \rho_q, \rho_{\bar q}$ stand for the
gluon, quark and antiquark spectral functions, respectively, and
$f_q$ $(f_{\bar q})$ stand for the equilibrium distribution
functions for particle and antiparticle. The medium-dependent
relaxation times $\bar \tau_{q,g} (T, \mu_q)$ in (\ref{equ:Sec5.1}) are
given in the DQPM$^*$ by:

{\setlength\arraycolsep{-1pt}
\begin{eqnarray}
\label{equ:Sec5.2} \bar \tau_{q,g} (T, \mu_q) = (\bar \gamma_{q,g})^{-1}
(T,\mu_q),
\end{eqnarray}}
with:
{\setlength\arraycolsep{-1pt}
\begin{eqnarray}
\label{equ:Sec5.3} & &  \bar \gamma_{q,g} (T, \mu_q) = \displaystyle \langle\gamma_{q,g} (T, \mu_q, p)\rangle_p = \left(n_{q,g}^{\textrm{off}} (T,\mu_q)\right)^{-1} \ \times \displaystyle \int \frac{d^3p}{(2 \pi)^3} \ \frac{d \omega}{(2 \pi)} \ \omega \ \gamma_{q,g} (T, \mu_q, p) \rho_f (\omega) \ f_{q,g} (\omega, T, \mu_q) \ \Theta(P^2),
\end{eqnarray}}
where $$n_{f,g}^{\textrm{off}} (T,\mu_q) = \displaystyle \int
\frac{d^3p}{(2 \pi)^3} \ \frac{d \omega}{(2 \pi)} \ \omega \ \rho_f
(\omega) \ f_{f,g} (\omega, T, \mu_q) \ \Theta(P^2)$$
denotes the
off-shell density of quarks, antiquarks or gluons. We note in
passing that the shear viscosity $\eta$ can also be computed using
the stress-energy tensor and the Green-Kubo formalism
\cite{Ozvenchuk:2012kh}. However, explicit comparisons of both
methods in Ref. \cite{Ozvenchuk:2012kh} have shown that the
solutions are rather close. This holds especially for the case of
the scattering of massive partons where the transport cross section
is not very different from the total cross section as also pointed
out in Ref. \cite{Plumari:2012ep}.

\begin{figure}[h!] 
\begin{center}
\begin{minipage}{20pc}
\includegraphics[width=19pc, height=18pc]{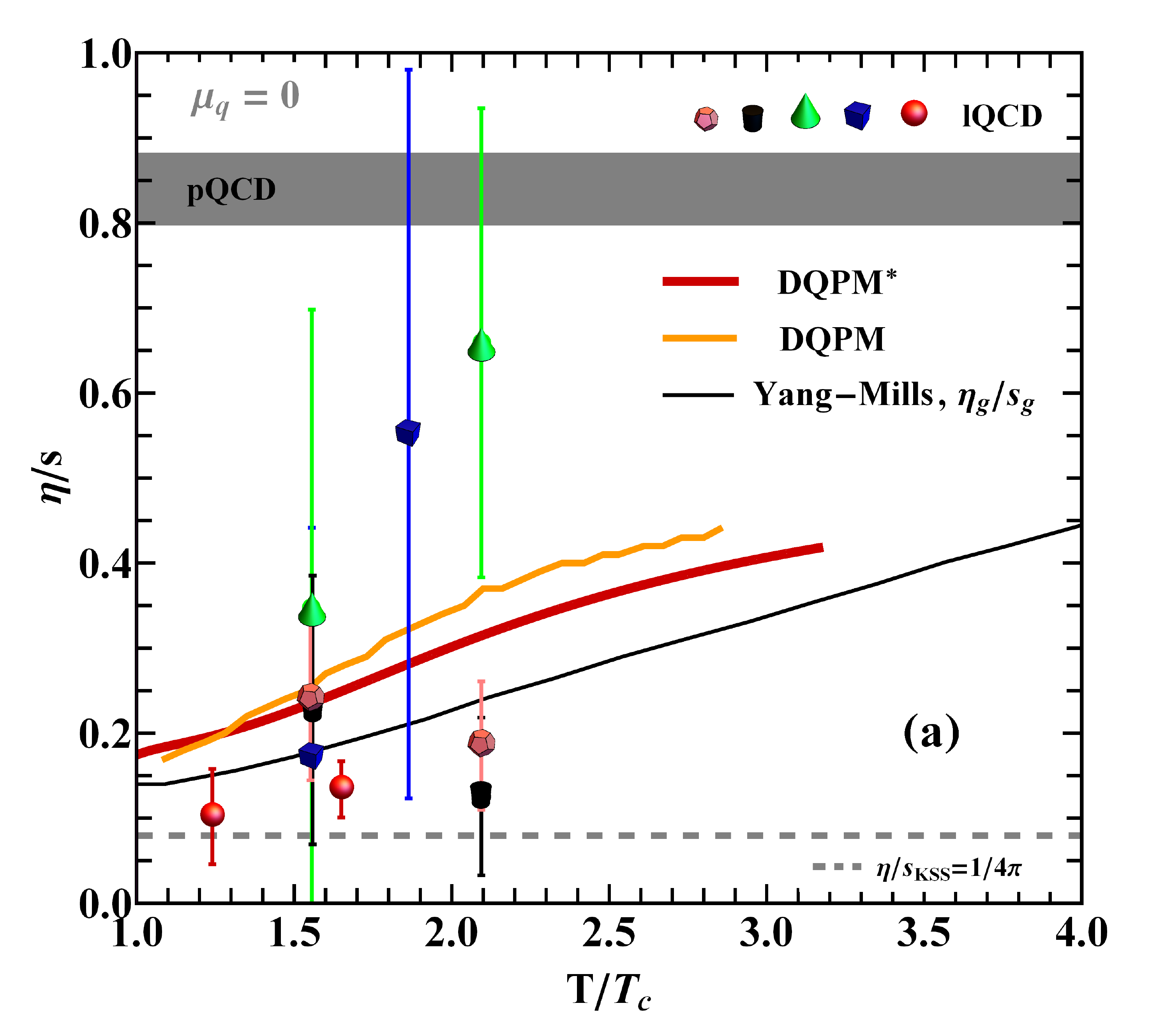}
\end{minipage}
\begin{minipage}{20pc}
\includegraphics[width=19pc, height=18pc]{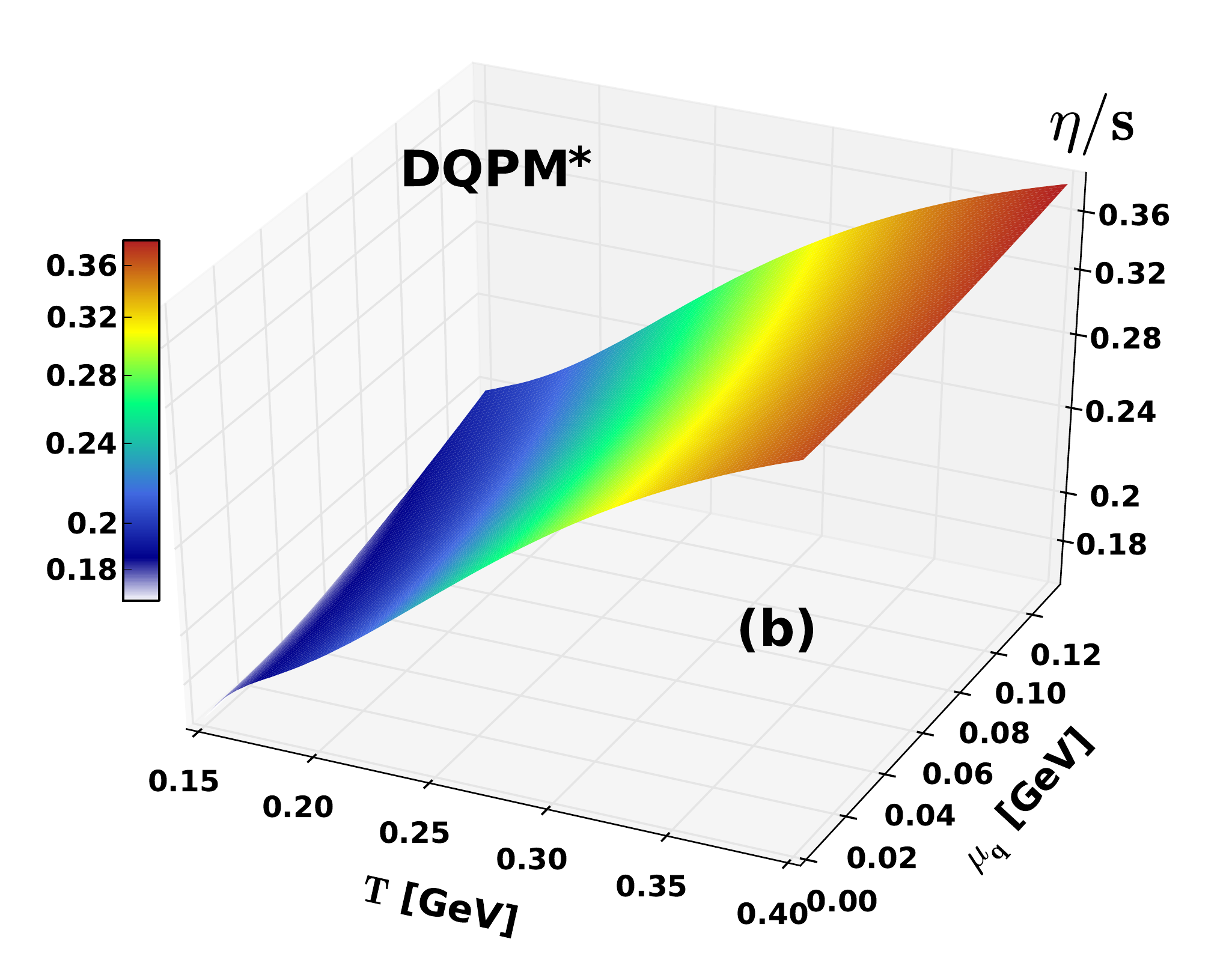}
\end{minipage}
\end{center}
\caption{\emph{(Color online) The shear viscosity to entropy density
ratio $\eta/s$ from different models as a function of temperature
$T$ for $\mu_q =0$ (a) and $\eta/s$ given by the pDQPM$^*$ approach
as a function of $(T, \mu_q)$ (b). The orange solid line in (a)
results from the standard DQPM where the parton masses and widths
are independent of momenta \cite{Berrehrah:2014ysa}. The thick red
solid line shows the DQPM$^*$ result using Eqs.(\ref{equ:Sec5.1})
and (\ref{equ:Sec5.2}), where the parton masses and width are
temperature, chemical potential and momentum dependent. The lattice
QCD data for pure $SU(3)$ gauge theory are taken from Ref.
\cite{Meyer:2007ic} (red spheres), from Ref. \cite{Nakamura:2004sy}
(green pyramid and blue cubic), and from Ref. \cite{Sakai:2007cm}
(black cylinder and pink pynthagone). The orange dashed line gives
the Kovtun-Son-Starinets lower bound
\cite{Policastro:2001yc,Kovtun:2004de} $(\eta/s)_{KSS} = 1/(4\pi)$.
Finally, the black solid line refers to the calculation of $\eta/s$
in Yang-Mills theory from Ref. \cite{Christiansen:2014ypa}.}}
\label{fig:EtaTmu}
\end{figure}

We show the DQPM$^*$ results for $\eta/s$, where $s$ is the DQPM$^*$
entropy density, in Fig.\ref{fig:EtaTmu} (a) as a function of the
temperature. The (upper) orange solid line represents the case of
the standard DQPM where the parton masses and widths are independent
of momenta as calculated in Ref. \cite{Berrehrah:2014ysa}. The thick
red solid line displays the result obtained in this study using
Eqs.(\ref{equ:Sec5.1}) and (\ref{equ:Sec5.2}), where the parton
masses and width are temperature, chemical potential and momentum
dependent. Finally, the black solid line refers to the calculation
of $\eta/s$ in Yang-Mills theory from the Kubo formula using an
exact diagrammatic representation in terms of full propagators and
vertices from Ref. \cite{Christiansen:2014ypa}.

Fig. \ref{fig:EtaTmu} (a) shows that $\eta/s$ from DQPM$^*$ is in
the range of the lQCD data and significantly lower than the pQCD
limit. As a function of temperature $\eta/s$ shows a minimum around
$T_c$, similar to atomic and molecular systems \cite{Csernai:2006zz}
and then increases slowly for higher temperatures. This behavior is
very much the same as in the standard DQPM (upper orange line) as
shown in Ref. \cite{Ozvenchuk:2012kh}. Therefore, the produced QGP
shows features of a strongly interacting fluid unlike a weakly
interacting parton gas as had been expected from perturbative QCD
(pQCD). The minimum of $\eta/s$ at $T_c = 158$ MeV is close to the
lower bound of a perfect fluid with $\eta/s = 1/(4 \pi)$
\cite{Policastro:2001yc,Kovtun:2004de} for infinitely coupled
supersymmetric Yang-Mills gauge theory (based on the AdS/CFT duality
conjecture). This suggests the ''hot QCD matter'' to be the ''most
perfect fluid'' \cite{Csernai:2006zz}. Furthermore, the ratio
$\eta/s$ in DQPM$^*$ is slightly larger than in the pure gluonic
system (solid black line) due to a lower interaction rate of quarks
relative to gluons.

The explicit dependencies of $\eta/s$ on $T$ and $\mu_q$ are shown
in Fig.\ref{fig:EtaTmu} (b) where $\eta/s$ is seen to increase
smoothly for finite but small $\mu_q$. We point out again that
extrapolations to larger $\mu_q$ become increasingly uncertain.

\section{Electric conductivity of the QGP from DQPM$^*$}
\label{etaOvs}

Whereas the shear viscosity $\eta$ depends on the properties of
quarks and gluons the electric conductivity $\sigma_e$ only depends
on quarks and antiquarks and thus provides independent information
on the response of the QGP to external electric fields
\cite{Cassing:2013iz,Steinert:2013fza}. The electric conductivity $\sigma_e$ is also
important for the creation of electromagnetic fields in
ultra-relativistic nucleus-nucleus collisions from partonic
degrees-of-freedom, since $\sigma_e$ specifies the imaginary part of
the electromagnetic (retarded) propagator and leads to an
exponential decay of the propagator in time $\sim \! \exp(-\sigma_e
(t-t'))$. Furthermore, $\sigma_e$ also controls the
photon spectrum in the long wavelength limit \cite{Linnyk:2015tha}.

We repeat here our previous studies on the electric conductivity
$\sigma_e(T)$ \cite{Cassing:2013iz,Marty:2013ita,Steinert:2013fza}
for `infinite parton matter' within the DQPM$^*$ using the novel
parametrizations of the dynamical degrees of freedom. We recall that
the dimensionless ratio $\sigma_e/T$ in the quasiparticle approach
is given by the relativistic Drude formula,
{\setlength\arraycolsep{-1pt}
\begin{eqnarray}
\label{equ:Sec6.1} & & \sigma_e (T,\mu_q) = \sum_{f,\bar{f}}^{u,d,s}
\frac{e_f^2 \ n_f^{\textrm{off}} (T, \mu_q)}{\bar \omega_f (T,
\mu_q) \ \bar \gamma_f (T, \mu_q)},
\nonumber\\
& & {} \textrm{with:} \  \bar \omega_f (T, \mu_q)=
\left(n_{f}^{\textrm{off}} (T,\mu_q)\right)^{-1} \ \times
\displaystyle \int \frac{d^3p}{(2 \pi)^3} \ \frac{d \omega}{(2 \pi)}
\ \omega^2 \  \rho_f (\omega, {\bs p}) \ f_f((\omega\pm \mu_q)/T),
\end{eqnarray}}
where the quantity $\bar \omega_q(T,\mu_q)$ is the quark (antiquark)
energy averaged over the equilibrium distributions at finite $T$ and
$\mu_q$ while $\bar \gamma_q(T,\mu_q)$ is the averaged quark width,
as given in (\ref{equ:Sec5.3}).

\begin{figure}[h!] 
\begin{center}
\begin{minipage}{20pc}
\includegraphics[width=19pc, height=18pc]{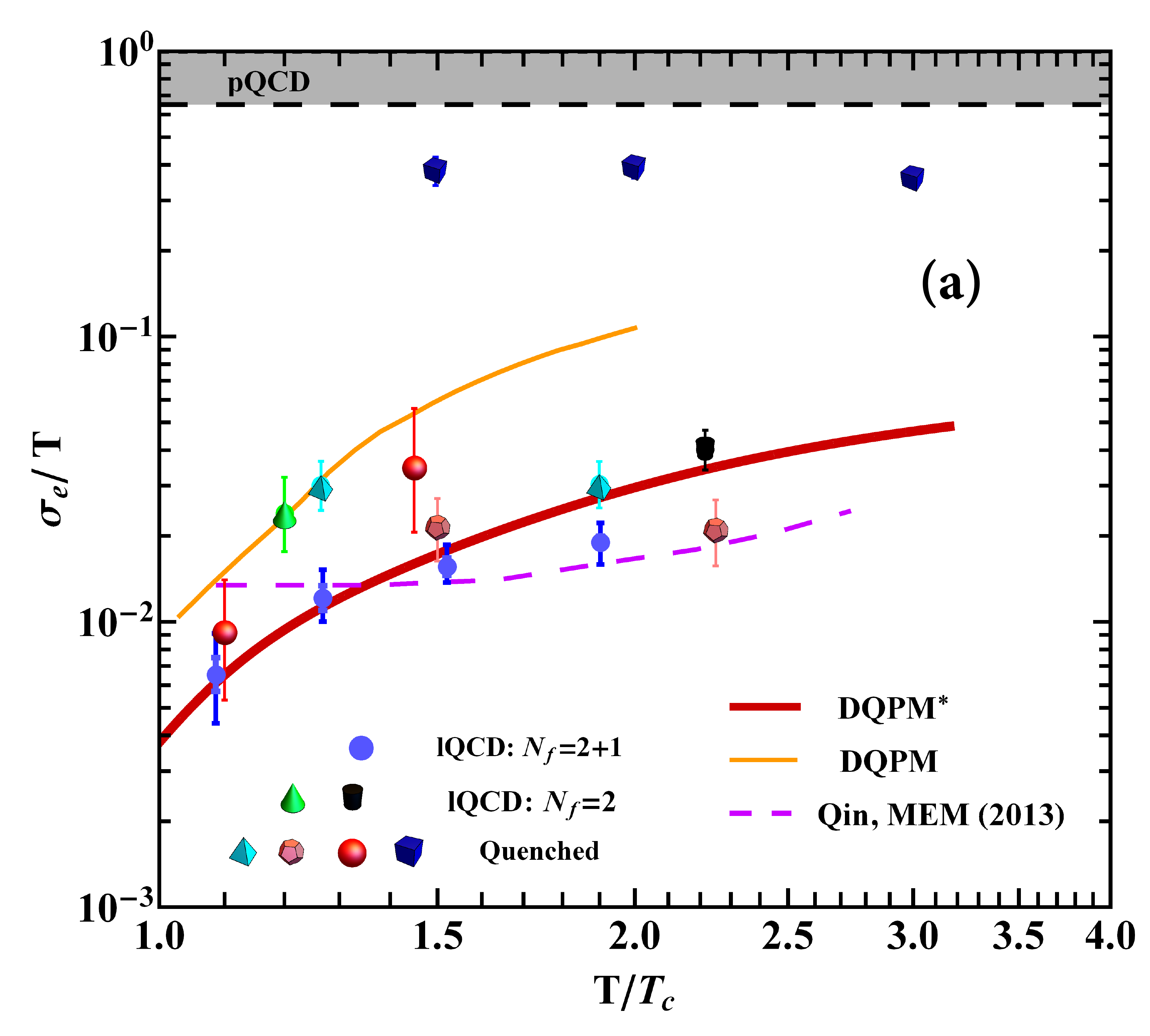}
\end{minipage}
\begin{minipage}{20pc}
\includegraphics[width=19pc, height=18pc]{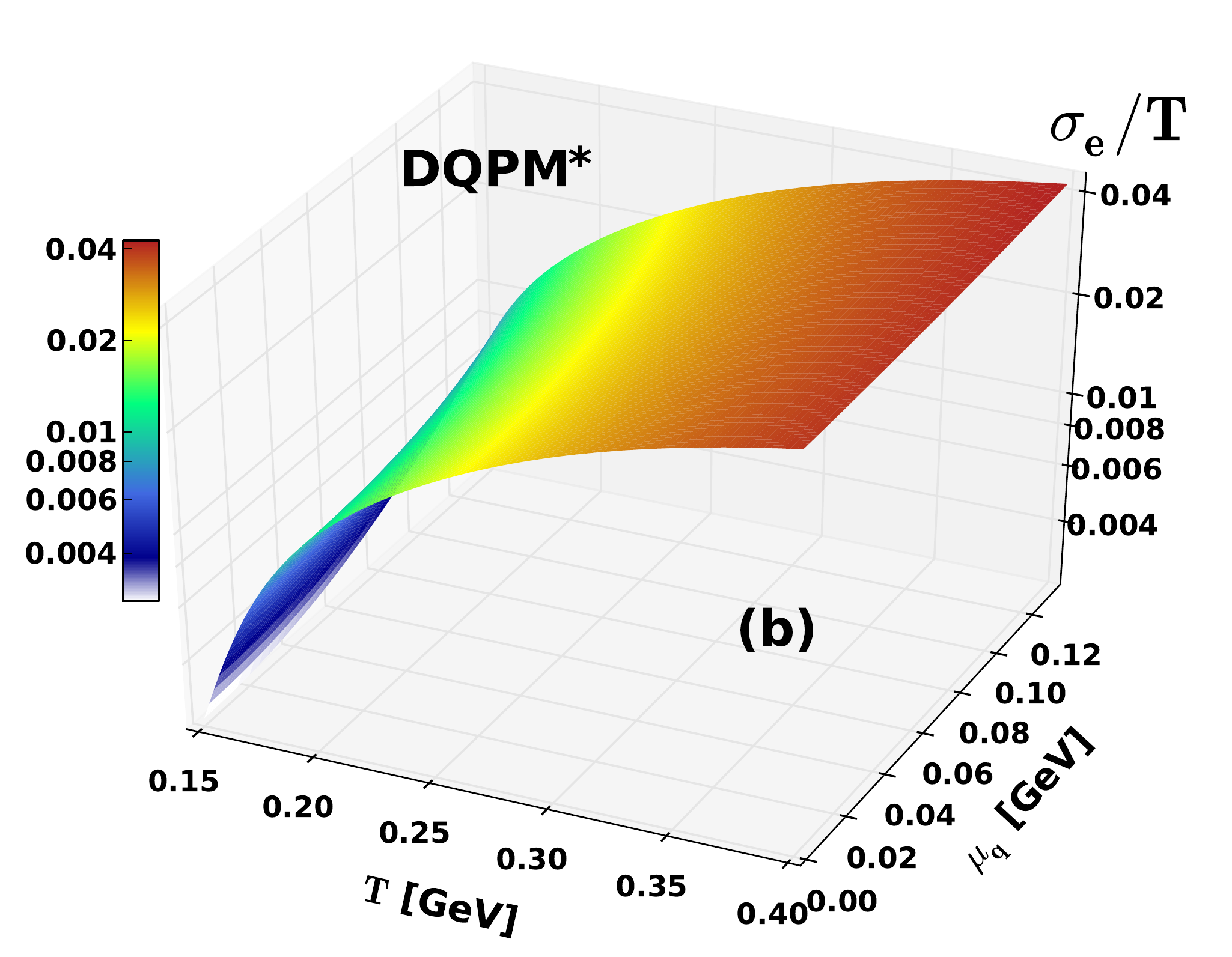}
\end{minipage}
\end{center}
\caption{\emph{(Color online) $\sigma_e/T$ following different
models as a function of temperature $T$ for $\mu_q =0$ (a) and
$\sigma_e/T$ given by the DQPM$^{\star}$ approach as a function of
$(T, \mu_q)$ (b). The orange thin solid line in (a) results from the
standard DQPM where the parton masses and widths are independent of
momenta \cite{Berrehrah:2014ysa}. The red thick solid line shows the
DQPM$^*$ result  using Eqs.(\ref{equ:Sec6.1}), where the parton
masses and width are temperature, chemical potential and momentum
dependent. The lattice QCD data are taken from Ref.
\cite{Ding:2010ga} (red spheres), Ref. \cite{Aarts:2007wj} (pink
pentagon), Ref. \cite{Gupta:2003zh} (blue cubic), Ref.
\cite{Kaczmarek:2013dya} (Cyan pyramid), Ref. \cite{Brandt:2013fg}
(green cone), Ref. \cite{Buividovich:2010tn} (black cylinder), Ref.
\cite{Aarts:2014nba} (blue disk). Qin, MEM (2013) refers to Ref.
\cite{Qin:2013aaa} where a Dyson-Schwinger approach is used. The
electric charge is explicitly multiplied out using $e^2 = 4
\pi/137$. The average charge squared is $C_{EM} = 8 \pi \alpha/3$
with $\alpha=1/137$. Note that the pQCD result at leading order
beyond the leading log \cite{Arnold:2003zc} is $\sigma_e/T \approx
5.97/e^2 \approx 65$.}} \label{fig:SigeTmu}
\end{figure}

The actual results for $\sigma_e/T$ are displayed in
Fig.\ref{fig:SigeTmu} (a) in terms of the thick red solid line in
comparison to recent lQCD data from Refs.
\cite{Ding:2010ga,Aarts:2007wj,Gupta:2003zh,Kaczmarek:2013dya,Brandt:2013fg,Buividovich:2010tn,Aarts:2014nba}
and the result from our previous studies within the DQPM
\cite{Marty:2013ita} (thin orange line). Again we find a minimum in
the partonic phase close to $T_c$ and a rise with the temperature
$T$. The explicit dependencies of $\sigma_e/T$ on $T$ and $\mu_q$,
shown in Fig.\ref{fig:SigeTmu} (b), is also increasing smoothly for
finite but small $\mu_q$. We finally note that the lower values for
$\sigma_e/T$ in the DQPM$^*$ relative to the DQPM result from using
the relativistic Drude formula (\ref{equ:Sec6.1}) instead of its
nonrelativistic counterpart.

\section{Summary}
\label{summary}

We have presented in this work an extension of the dynamical
quasiparticle model (DQPM) with respect to momentum-dependent
selfenergies in the parton propagators which are reflected in
momentum-dependent masses and widths. Accordingly, the QGP effective
degrees of freedom appear as interacting off-shell quasi-particles
with masses and widths that depend on momentum $p$, temperature $T$
and chemical potential $\mu_q$ as given in Eqs. (\ref{equ:Sec2.1}).
These expressions provide a proper high temperature limit as in the
HTL approximation and approach the pQCD limit for large momenta $p$.
As in the standard DQPM the effective coupling is enhanced in the
region close to $T_c$ which leads to an increase of the parton
masses roughly below 1.2 $T_c$ (cf. Fig. 1 a)). Instead of
displaying the parton masses as a function of temperature we may
alternatively display them as a function of the scalar parton
density $\rho_s$ (cf. Ref. \cite{Bratkovskaya2011162}) and interpret
the masses as a scalar mean-field depending on $\rho_s$. Since
$\rho_s$ is a monotonically increasing function with temperature the
masses $M_j(\rho_s)$ will show a minimum in $\rho_s$ for $\rho_s
\approx $ 0.5 fm$^{-3}$ which specifies the parton density where the
effective interaction -- defined by the derivative of the masses
with respect to the scalar density -- changes sign, i.e. the net
repulsive interaction at high scalar density becomes attractive at
low scalar density and ultimately leads to bound states of the
constituents (cf. Ref. \cite{Cassing2007365}).

The extended dynamical quasiparticle model is denoted by DQPM$^*$
and reproduces quite well the lQCD results, i.e. the QGP equation of
state, the baryon density $n_B$ and the quark susceptibility
$\chi_q$  at finite temperature $T$ and quark chemical potential
$\mu_q$  which had been a challenge for quasiparticle models so far
\cite{Plumari:2011mk} (see also Fig. 3b). A detailed comparison
between the available lattice data and DQPM$^*$ results indicates a
very good agreement for temperatures above $\sim$ 1.2 $T_c$ in the
pure partonic phase and therefore validates our description of the
QGP thermodynamic properties. For temperatures in the vicinity of
$T_c$ (and $\mu_B$= 400 MeV) we cannot expect our model to work so
well since here hadronic degrees of freedom mix in a crossover phase
which are discarded in the DQPM$^*$.

Furthermore, we have computed also the QGP shear viscosity $\eta$
and electric conductivity $\sigma_e$ at finite temperature and
chemical potential in order to probe some transport properties of
the medium. The relaxation times at finite temperature and chemical
potential, used in our study, are evaluated for the dynamical
quasi-particles using the parton width which is averaged over the
thermal ensemble at fixed $T$ and $\mu_q$. We, furthermore,
emphasize the importance of nonperturbative effects near $T_c$ to
achieve a small $\eta/s$ as supported by different phenomenological
studies and indirect experimental observations. When comparing our
results for $\eta/s$ to those from the standard DQPM in Ref.
\cite{Ozvenchuk:2012kh} we find a close agreement. In the DQPM$^*$
the gluon mass is slightly higher (for low momenta)  and the quark
mass is slightly smaller than in the DQPM. Furthermore, the
interaction widths are slightly larger in the DQPM$^*$ which finally
leads to a slightly lower shear viscosity $\eta$ than in the DQPM.
This also holds for the electric conductivity $\sigma_e$ which in
the DQPM$^*$ gives results even closer to the present lQCD 'data'.

In view of our results on the description of QGP thermodynamics and
transport properties, one can conclude that the DQPM$^*$ provides a
promising approach to study the QGP in equilibrium at finite
temperature $T$ and chemical potential $\mu_q$. Moreover, we have
demonstrated, for the first time, that one can simultaneously
reproduce the lQCD pressure and quark susceptibility using a
dynamical quasi-particle picture for the QGP effective degrees of
freedom. An implementation of the DQPM$^*$ in the PHSD transport
approach \cite{Bratkovskaya2011162} is straight forward and will
allow for the description of heavy-ion collisions also for invariant
energies $\sqrt{s_{NN}} \approx$ 5-10 GeV.

\section{acknowledgment}
\label{acknowledgment}

This work has been supported by the ``HIC for FAIR'' framework of
the ``LOEWE'' program. The computational resources have been
provided by the LOEWE-CSC. Furthermore, the authors acknowledge
valuable discussions with R. Marty, P.B. Gossiaux, J. Aichelin, P.
Moreau and E. Seifert.

\bibliography{References}

\end{document}